\magnification=1200
\def\b{\beta}
\def\h{\theta}
\def\k{\kappa}\def\l{\lambda}\def\m{\mu}\def\n{\nu}\def\r{\rho}\def\s{\sigma}
\def\y{\eta}\def\x{\xi}\def\z{\zeta}

\def\de{\partial}
\def\id{\equiv}\def\mo{{-1}}

\def\({\left(}\def\){\right)}\def\[{\left[}\def\]{\right]}
\def\bdot{\!\cdot\!}

\def\arcth{\mathop{\rm arctanh}\nolimits}
\def\arcsh{\mathop{\rm arcsinh}\nolimits}

\def\mn{{\mu\nu}}

\def\tran{transformations }\def\coo{coordinates }

\def\rep{representation }

\def\pb{Poisson brackets }

\def\poi{Poincar\'e }

\def\eom{equations of motion }

\def\sys{symplectic structure }

\def\section#1{\bigskip\noindent{\bf#1}\smallskip}

\def\PL#1{Phys.\ Lett.\ {\bf#1}}
\def\PRL#1{Phys.\ Rev.\ Lett.\ {\bf#1}}
\def\PR#1{Phys.\ Rev.\ {\bf#1}}\def\CQG#1{Class.\ Quantum Grav.\ {\bf#1}}

 \def\IJMP#1{Int.\ J. Mod.\ Phys.\ {\bf #1}}

\def\AoP#1{Ann.\ Phys.\ {\bf#1}}

\def\JHEP#1{JHEP\ {\bf#1}}\def\JCAP#1{JCAP\ {\bf#1}}

\def\arx#1{{\tt arXiv:#1}}

\def\ref#1{\medskip\everypar={\hangindent 2\parindent}#1}
\def\beginref{\begingroup
\bigskip
\centerline{\bf References}
\nobreak\noindent}
\def\endref{\par\endgroup}

\def\rs{{\r\s}}\def\cA{{\cal A}}\def\cB{{\cal B}}

\magnification=1200

{\nopagenumbers
\line{}
\vskip30pt
\centerline{\bf Relative-locality effects in Snyder spacetime}

\vskip60pt
\centerline{
{\bf S. Mignemi}$^{1,2,}$\footnote{$^\ddagger$}{e-mail: smignemi@unica.it},
and {\bf A. Samsarov}$^{3,}$\footnote{$^*$}{e-mail: samsarov@unica.it},}
\vskip10pt
\centerline{$^1$Dipartimento di Matematica e Informatica, Universit\`a di Cagliari}
\centerline{viale Merello 92, 09123 Cagliari, Italy}
\smallskip
\centerline{$^2$INFN, Sezione di Cagliari, Cittadella Universitaria, 09042 Monserrato, Italy}
\smallskip
\centerline{$^3$Rudjer Bo\v skovi\'c Institute, Bijeni\v cka cesta 54, 10002 Zagreb, Croatia}
\vskip80pt

\centerline{\bf Abstract}
\medskip
{\noindent
When applied to some models of noncommutative geometry, the formalism of relative locality predicts
the occurrence of a delay in the time of arrival of massless particle of different energies emitted
by a distant observer. In this letter, we show that this is not the case with Snyder spacetime,
essentially because the Lorentz invariance is not deformed in this case.
This conclusion is in accordance with the findings of doubly special relativity.
Distant observers however may measure different times of flight for massive particle.}
\vskip10pt
{\noindent

}
\vskip80pt\
\vfil\eject}
\section{1. Introduction}
Recently, relative locality [1] has been proposed as a framework for investigating the physical
properties of models derived from noncommutative geometry (NCG) [2] and doubly special relativity (DSR) [3],
and hence giving an effective description of quantum gravity effects in the limit were $\hbar$
and $G$ become negligible, but their ratio $M_P=\sqrt{\hbar/G}$ stays finite,
providing a fundamental energy scale that can deform the kinematics of special relativity.

The formalism of relative locality is based on the postulate that momentum space is curved
and the nontrivial
physical effects arising from NCG and DSR (such as deformed dispersion relations,
noncommutativity and nonassociativity of the addition of momenta, etc.)
are related to nontrivial properties of the geometry
of momentum space, as torsion, curvature, and nonmetricity.
The main physical implication of the nontrivial geometry of momentum space consists in the loss of
the absolute meaning of the concept of locality, that becomes observer dependent.
The theory proved to be very well suited in particular for the
description of models based on the noncommutative $\k$-Minkowski spacetime [4].

One of the most striking consequences of relative locality for phenomenology is the
possibility that the speed of massless particles can depend on their energy, in accordance
with some predictions of NCG [5] and DSR (although in the latter case this result depends on the
details of the dynamics and on the identification of the velocity of particles with the group
velocity ${\bf v}_i=\de p_0/\de p_i$ rather than with the kinematic definition
${\bf v}_i=\dot x_i/\dot x_0$ [6]).
This effect would imply a time delay in the observation of particles of different
energy simultaneously emitted from a distant observer.
In particular, in [7] it was shown that in the framework of relative locality this result
can be deduced from a purely classical
analysis of the wordlines of simultaneously emitted particles of different momenta and from
the relativity of that simultaneity for distant observers. The proof does not rely on a specific
definition of velocity, and was extended to more general models in [8].

It is interesting to investigate if relative locality predicts an energy dependence of the speed
of light also in the case of the Snyder model [9].
This is a model of noncommutative geometry whose distinctive feature is the preservation of the \poi
invariance. Therefore, the dispersion relation for particles is essentially the same as in special
relativity, and the speed of light should be independent of the momentum of the particles.
More explicitly, the dispersion relation is a function of $p^2=p_0^2-p_i^2$.
It follows that both definitions of velocity mentioned above give the same result,
${\bf v}_i=p_i/p_0$, and in particular massless particles move at the speed of light.
Consistency of the relative locality formalism would therefore require
that using arguments analogous to those of [7] one can predict for Snyder spacetime
a momentum-independent speed of light
and hence no delay in the detection of simultaneously emitted particles of different energy.
The purpose of this letter is to show that it is indeed so.

To obtain this result, we discuss the geometry of the momentum space
of the Snyder model, and establish the correct Hamiltonian according to the prescriptions of
relative locality, solving the ambiguity related to the fact that, in the context of DSR theories, any
function of the Casimir
invariant of the \poi algebra can be chosen as Hamiltonian for the Snyder model. We show that the \eom for a
free particle are equivalent to those of special relativity, giving rise to the same geodesic motion.
The equivalence of course disappears in the case of an external coupling.

\section{2. The geometry of the Snyder model}
The Snyder model is defined by the \pb [9]
$$\eqalign{&\{J_\mn,J_\rs\}=\y_{\m\r}J_{\n\s}-\y_{\m\s}J_{\n\r}-\y_{\n\r}J_{\m\s}+\y_{\n\s}J_{\m\r},\cr
&\quad\{J_\mn,p_\r\}=\y_{\m\r}p_\n-\y_{\n\r}p_\m,\qquad\{p_\m,p_\n\}=0,}\eqno(1)$$
$$\{J_\mn,x_\r\}=\y_{\m\r}x_\n-\y_{\n\r}x_\m,\qquad\{x_\m,p_\n\}=\y_\mn-\b^2p_\m p_\n,
\qquad\{x_\m,x_\n\}=-\b^2J_\mn,\eqno(2)$$
where $\b$ is a constant of order $1/M_P$, $\m,\n=0,\dots,3$, $\y_\mn=$ diag $(1,-1,-1,-1)$ is the flat
metric and $J_\mn=x_\m p_\n-x_\n p_\m$ are the generators of the Lorentz transformations.
Eqs.\ (1) reproduce the \poi algebra, while eqs.\ (2) describe the action of the \poi
group on the position coordinates and the noncommutativity of spacetime.

Since the \poi algebra is not deformed, the momenta transform in the standard way, in
particular are unaffected by translations. For what concerns the position coordinates,
the Lorentz \tran act on them in the usual way, while the action of the translations is nontrivial.
In fact, from (2) it follows that under a finite
translation of parameter $a_\m$,\footnote{$^1$}{In the following we denote
$A\bdot B=\y^\mn A_\m B_\n$, $A^2=A\bdot A$ and $|A|=\sqrt{A^2}$.}
$$x_\m\to x_\m+a_\m-\b^2a\bdot p\ p_\m.\eqno(3)$$
Hence, the effect of a translation depends on the 4-momentum of the particle, as in most examples of
DSR and NCG [10,11].

As other models of NCG and DSR related to relative locality,
the Snyder model can be realized on a constant curvature momentum space [12,13].
More precisely, the momentum space of the Snyder model can be identified with a hyperboloid
embedded in a five-dimensional flat space of
\coo $\x_A$ $(A=0,\dots,4)$, with signature $(1,-1,-1,-1,-1)$, satisfying the constraint
$\x_A^2=-1/\b^2$.
Since Lorentz invariance is preserved,
it is convenient to parametrize the space using isotropic coordinates.
This can be done in several different ways, which are however not equivalent.

The simplest parametrization is obtained by identifying the four-dimensional momentum $p_\m$ as
 $p_\m=\x_\m$. The metric induced on the hyperboloid and its inverse are then
$$g_\mn=\y_\mn-{\b^2p_\m p_\n\over1+\b^2p^2},\qquad\qquad g^\mn=\y^\mn+\b^2p^\m p^\n,\eqno(4)$$
where $\y_\mn$ is the four-dimensional Minkowski metric. This parametrization holds for
$p^2>-1/\b^2$ and is therefore compatible with any value of mass.

Another remarkable possibility are the so-called Beltrami coordinates, defined as
$\b p_\m=\x_\m/\x_4$, with metrics
$$g_\mn={(1-\b^2p^2)\y_\mn+\b^2p_\m p_\n\over(1-\b^2p^2)^2},\qquad\qquad
g^\mn=(1-\b^2p^2)(\y^\mn-\b^2p^\m p^\n).\eqno(5)$$
In this case, the inverse transformations are $\x_\m=p_\m/\sqrt{1-\b^2p^2}$,
$\b\x_4=1/\sqrt{1-\b^2p^2}$. From the definition follows that the bound $p^2<1/\b^2$ must
be satisfied; hence an upper limit exists on the mass of particles.
The existence of bounds on the mass or on the energy of particles is a common feature in DSR theories [10].

To complete the definition of the Snyder phase space, one must choose the position \coo
$x_\m$ in such a way that they satisfy the \pb (2). Calling $\z_A$ the five-dimensional
position \coo
canonically conjugated to the $\x_A$, so that $\{\z_A,\x_B\}=\y_{AB}$, one can show that
in the representation (4) the position \coo are given by $x_\m=\z_\m-\b^2\z_A\x^A\x_\m$, while
in the Beltrami \rep they are given by  $x_\m=\b(\x_4\z_\m-\z_4\x_\m)$ and coincide with the
generators of translations in the hyperboloid of momenta.

Due to the nontrivial symplectic structure, the classical dynamics of the Snyder model is
suitably described in Hamiltonian form.
The Hamiltonian can be an arbitrary function of the Casimir operator $p^2$
of the undeformed \poi algebra.
To fix this arbitrariness, in relative locality models the Hamiltonian is defined as the
square of the
geodesic distance in momentum space from the origin to a point parametrized by $p_\m$ [1].
For isotropic \coo this
is very easily computed geometrically, without resorting to the calculation of the geodesics.
On a hyperboloid, the distance from the origin along a timelike path is given simply by
$$l={1\over\b}\arcth{|\x|\over\x_4}={1\over\b}\arcsh\b|\x|.\eqno(6)$$
This can be shown considering the embedding of the Snyder hyperboloid in five-dimensional Minkowski space.
Because of the isotropy, one can simply consider a one-dimensional section in two-dimensional embedding
space. For timelike geodesics, this is a hyperbola, that can be parametrized by $\b\x_0=\sinh\h$,
$\b\x_1=\cosh\h$. The arclength is then
${l=\int\sqrt{\dot \x_0^2-\dot \x_1^2}\ d\h=\b^\mo\int\sqrt{\cosh^2\h-\sinh^2\h}\ d\h=\b^\mo\h}$,
from which (6) readily follows.

Hence the Hamiltonian for a free particle of mass $m$ in the \coo (4) is
$$H={\l\over2}\({\arcsh^2\b|p|\over\b^2}-m^2\),\eqno(7)$$
while in Beltrami \coo it is
$$H={\l\over2}\({\arcth^2\b|p|\over\b^2}-m^2\),\eqno(8)$$
with $\l$ a Lagrange multiplier enforcing the Hamiltonian constraint.
Notice that in the previous literature [14] the Hamiltonian has been chosen to be $p^2$ or some
simple algebraic function of it, as $p^2/(1-\b^2p^2)$. Although in the case of free particles
the geodesics are not modified, except for a reparametrization of the momentum, in
the interacting case important effects may arise.

Of particular interest are the \eom that follow from the Hamiltonian (8) and the \sys (2),
$$\dot x_\m=\l q_\m\id\l {\arcth\b|p|\over\b|p|}\ p_\m,\qquad\dot p_\m=\dot q_\m=0.\eqno(9)$$
The geodesic equations for a free particle are identical to those of special relativity
when written in terms of the auxiliary variable $q_\m$ if $m\ne0$ and coincide with them if $m=0$.
It follows that the momentum components are conserved and the geodesics are then given by
$$x_i=\bar x_i+{\bar q_i\over \bar q_0}(x_0-\bar x_0)=\bar x_i+{\bar p_i\over \bar p_0}(x_0-\bar x_0),\eqno(10)$$
 which are exactly the same as in special relativity.

The Hamiltonian (7) yields slightly more involved equations,
$$\dot x_\m=\l r_\m\id\l {\sqrt{1+\b^2p^2}\over1-\b^2p^2}\ {\arcsh\b|p|\over\b|p|}\ p_\m,
\qquad\dot p_\m=\dot r_\m=0,\eqno(11)$$
but the geodesics are still given by (10).

\section{3. Relative locality effects}
Following [7], we now use arguments from classical relativistic mechanics and relative locality
to evaluate the delay in the time of arrival of massless particles of different
momenta emitted simultaneously according to an observer $\cA$, as detected by a distant observer
$\cB$ in Snyder space.
It has been shown in [7] that such effect is present in $\k$-Minkowski spaces, giving rise to the
possibility of detecting experimentally signals of the structure of spacetime at Planck scale.

From the Lorentz invariance of the model and its analysis in terms of DSR, we expect that this effect should
not arise in the case of Snyder spacetime. However, the nontrivial \tran of spacetime under translations (3)
might invalidate this conclusion and it is therefore necessary to analyze in detail this topic.

For simplicity we consider a two-dimensional  setting,
and denote $x_0=t$, $x_1=x$, $p_0=E$, $p_1=P$.
Without loss of generality we can choose the initial conditions at an event $\cA$
with $\bar x=0$, $\bar t=0$.

We suppose that at $\cal A$ two massless particles are emitted with momenta $P_1=E_1$ and $P_2=E_2$,
according to the dispersion relation (8). In conformity with (10), their worldlines are given by
$$x_{P_1}^\cA(t^\cA)=t^\cA,\qquad x_{P_2}^\cA(t^\cA)=t^\cA,\eqno(13)$$
and of course, according to $\cA$, they reach $\cB$ at the same instant.

The distant observer $\cB$ who detects the particles is related to $\cA$ by a translation
of parameter $a_\m=(a,a)$, and therefore the quantities he measures are related to those measured by $\cA$
by the transformations (3).
As we have seen, in Snyder space the momenta $p_\m$ are invariant under translations,
while $x$ and $t$, according to (3), transform as
$$x^\cB=x^\cA+a,\qquad t^\cB=t^\cA+a,\eqno(14)$$
since $a\bdot p=a(E-P)$ vanishes in our setting.
Substituting in (13), it follows that
$$x_{P_1}^\cB(t^\cB)=x_{P_2}^\cB(t^\cB)=t^\cB,\eqno(15)$$
independently of the value of $P$.
Hence, as expected, the two particles will be detected by $\cB$ at the same time in contrast with other
noncommutative models, where a time delay arises [7,8].

In principle, it can be interesting to consider also the case of massive particles, although an experimental
verification does not seem to be realizable in this case.
For massive particles in Beltrami parametrization, eq.\ (8) yields $E=\sqrt{P^2+\b^{-2}\tanh^2\b m}$.
According to $\cA$, after a time $t^\cA=a$, the particles are in
a position $x^\cA={P\over E}\,a$. If the observer $\cB$ is placed in this position, he is related to $\cA$ by a
translation with parameter $a_\m=(a,{P\over E}\,a)$, and hence due to (3) he measures,
$$t^\cB=t^\cA+{a\over\cosh^2\b m},\qquad x^\cB=x^\cA+{a\over\cosh^2\b m}\ {P\over E},\eqno(16)$$
since $a\bdot p=a(E^2-P^2)/E=a\tanh^2\b m/\b^2E$.

Therefore,
$$x^\cB={P\over E}\ t^\cB,\eqno(17)$$
which is the same relation valid in special relativity. Hence, also massive Snyder particles do not present time delay
effects except for the trivial ones due to the different velocity. However, a difference arises in the time of flight
of the particles as
seen from the observer $\cB$ with respect to the observer $\cA$. In fact, they differ by a factor $1/\cosh^2\b m$;
this could induce anomalies in the lifetime of unstable massive particles measured by $\cB$, analogous to those
investigated in [15] in a different setting.

As we have seen, the geodesics do not depend essentially on the coordinates chosen.  Therefore the results of
this section are valid also in the case of the parametrization (4).

\section{4. Conclusions}
Using the formalism of relative locality,
we have shown that in the case of Snyder geometry no effect due to delayed time of arrival of high-energy
massless particles is observable, contrary to other models of noncommutative geometry [7,8].
This may have been expected because of the Lorentz invariance of the model
 and confirms the results one would get from a DSR approach.

However, different times of flight are measured by distant observers for massive particles of different
energies, and hence some effects can in principle be detected due to the energy dependence of the lifetime
of unstable particles.

It seems therefore that the relative locality effects are greatly mitigated if the Lorentz invariance is
not deformed. It is likely that this conclusion can be extended to more general Lorentz-invariant NCG models,
like those studied in refs.\ [16].

For the derivation of the results of this paper, only the metric of the Snyder momentum space was needed.
However, in the framework of relative locality one can associate also torsion and nonmetricity to the momentum
manifold, depending on the properties of the law of addition of the momenta [1]. We plan to discuss in detail
this topic in a future paper.
\bigskip
\section{Acknowledgments}
The research leading to these results has received funding from the European Union Seventh Framework Programme
(FP7 2007-2013) under grant agreement n 291823 Marie Curie FP7-PEOPLE-2011-COFUND NEWFELPRO as a
part of a project NCGGBH which has received funding through NEWFELPRO project under grant agreement n 63.
The work by A.S. has also been supported by Croatian Science Foundation under the project (IP-2014-09-9582).

\beginref

\ref [1] G. Amelino-Camelia, L. Freidel, J. Kowalski-Glikman and L. Smolin, \PR{D84}, 084010 (2011);
\CQG{29}, 075007 (2012).
\ref [2] S. Doplicher, K. Fredenhagen and J.E. Roberts, \PL{B331}, 39 (1994).
\ref [3] G. Amelino-Camelia, \PL{B510}, 255 (2001), \IJMP{D11}, 35 (2002).
\ref [4] G. Gubitosi and F. Mercati, \CQG{20}, 145002 (2013).
\ref [5]  G. Amelino-Camelia and S. Majid, \IJMP{A15}, 4301 (2000).
\ref [6] J. Lukierski and A. Nowicki, Acta Phys. Pol. {\bf B33}, 2537 (2002);
S. Mignemi, \PL{A316}, 173 (2003); P. Kosi\'nski and P. Ma\'slanska, \PR{D68}, 067702 (2003);
G. Amelino-Camelia, F. d'Andrea and G. Mandanici, \JCAP{0309}, 006 (2003);
M. Daszkiewicz, K. Imilkowska and J. Kowalski-Glikman, \PL{A323}, 345 (2004).
\ref [7] G. Amelino-Camelia, N. Loret and G. Rosati, \PL{B700}, 150 (2011).
\ref [8] S. Meljanac, A. Pachol, A. Samsarov and K.S. Gupta, \PR{D87}, 125009 (2013).
\ref [9] H.S. Snyder, \PR{71}, 38 (1947).
\ref [10] G. Amelino-Camelia, \IJMP{D11}, 1643 (2002); J. Magueijo and L. Smolin, \PRL{88}, 190403 (2002).
\ref [11] J. Lukierski, H. Ruegg and W.J. Zakrzewski, \AoP{243}, 90 (1995).
\ref [12] J. Kowalski-Glikman, \PL{B547}, 291 (2002);
J. Kowalski-Glikman and S. Nowak, \IJMP{D13}, 299 (2003).
\ref [13] F. Girelli and E. Livine, \JHEP{1103}, 132 (2011).
\ref [14] S. Mignemi, \PL{B672}, 186 (2002); \IJMP{D24}, 1550043 (2015);
R. Banerjee, S. Kulkarni and S. Samanta, \JHEP{0605}, 077 (2006);
F. Girelli, T. Konopka, J. Kowalski-Glikman and E.R. Livine, \PR{D73}, 045009 (2006).
\ref [15] S. Mignemi, \PL{A373}, 4401 (2009).
\ref [16] M.V. Battisti and S. Meljanac, \PR{D79}, 067505 (2009);
S. Meljanac, D. Meljanac, S. Mignemi and R. \v Strajn, \arx{1608.06207}.

\endref

\end